\documentclass[twocolumn]{aastex701}

\graphicspath{{./}{Plots/}{PRL/Plots/}}
\usepackage[normalem]{ulem}
\usepackage{dcolumn}   
\usepackage{bm} 
\usepackage[percent]{overpic}
\usepackage{ragged2e}   
\usepackage[english]{babel}

\usepackage{subcaption}

\usepackage{orcidlink}
\usepackage{soul}

\newcommand{\radeffins}{0.23}     
\newcommand{\radeffpost}{0.6}    
\newcommand{\dtpost}{5.9\times10^{3}}  

\begin{document}

\title{Accretion, Jets, and Recoil in a Merging Supermassive Black Hole Binary:
A Prompt Electromagnetic Postmerger Counterpart for LISA}



\correspondingauthor{Maria Chiara de Simone}

\author{Maria Chiara de Simone\,\orcidlink{0009-0008-8088-1392}}
\affiliation{Center for Computational Relativity and Gravitation, Rochester Institute of Technology \& School of Physics and Astronomy, Rochester, New York 14623, USA}
\email{md1556@rit.edu}

\author{Manuela Campanelli\,\orcidlink{0000-0002-8659-6591}}
\affiliation{Center for Computational Relativity and Gravitation, Rochester Institute of Technology \& School of Physics and Astronomy, Rochester, New York 14623, USA}
\email{manuela@astro.rit.edu}

\author{Lorenzo Ennoggi\,\orcidlink{0000-0002-2771-5765}}
\affiliation{Center for Computational Relativity and Gravitation, Rochester Institute of Technology \& School of Physics and Astronomy, Rochester, New York 14623, USA}
\email{le8016@rit.edu}

\author{Carlos O. Lousto\,\orcidlink{0000-0002-6400-9640}}
\affiliation{Center for Computational Relativity and Gravitation, Rochester Institute of Technology \& School of Physics and Astronomy, Rochester, New York 14623, USA}
\email{colsma@rit.edu}

\author{Yosef Zlochower\,\orcidlink{0000-0002-7541-6612}}
\affiliation{Center for Computational Relativity and Gravitation, Rochester Institute of Technology \& School of Mathematics and Statistics, Rochester, New York 14623, USA}
\email{yrzsma@rit.edu}

\begin{abstract}

We report the first three-dimensional general relativistic magnetohydrodynamic
simulation to \mbox{follow}, self-consistently in a dynamical spacetime, the
magnetized gas around a misaligned-spin supermassive binary black hole, from
late inspiral through merger to the gravitational-wave recoil of the remnant.
The equal-mass binary, in a \textit{hang-up kick} configuration, is embedded in
an equilibrated circumbinary disk (CBD), relaxed for 165 binary orbits before we
evolve the final $\sim\!40$ orbits. During the inspiral, each black hole hosts a
strongly warped minidisk whose jet follows the local spin axis near the horizon
before aligning with the binary's angular momentum farther out. The merger
imparts a recoil of $\simeq\!1032$ km\,s$^{-1}$ to the remnant, which
nonetheless retains its gravitationally bound CBD, and the relaunched jet
preserves its pre-merger orientation. The bolometric luminosity brightens by a
factor $\sim\!2.2$, powered by merger-driven shocks concentrated within
$r \lesssim 15\,M$, and the enhancement persists through the recoil. We identify
a distinctive postmerger electromagnetic signature: magnetized structures,
generated at coalescence, drive correlated quasi-periodic modulations of the
horizon magnetic flux, the Poynting flux, and the thermal output, decoupled from
the accretion rate. The postmerger radiative efficiency $L_{\rm EM}/\dot{M}c^2$
rises by a factor $\simeq 2.6$ at nearly constant accretion rate, showing that
this emission is powered by the merger rather than accretion.
The transient turns on within minutes and lasts at least several hours for a
$10^6 \ M_{\odot}$ LISA source, establishing recoiling remnants as prompt
postmerger counterparts to massive black hole mergers and a first-principles
framework for interpreting candidates such as 3C186.

\end{abstract}


\section{Introduction}

Supermassive binary black holes (SMBBHs) are prime targets for multi-messenger astrophysics. In the final stages of their evolution they radiate gravitational-wave (GW) power comparable to the combined luminosity of all stars in the observable Universe, and, when \mbox{embedded} in gas, the late inspiral, merger, and postmerger relaxation can each produce powerful electromagnetic (EM) emission. Pulsar timing arrays have recently detected a nanohertz GW background consistent with a large population of SMBBHs \citep{Agazie:2023,Antoniadis:2023}, while deep, wide, rapid-cadence surveys such as those of the Vera C. Rubin Observatory are poised to discover transients associated with individual systems. Identifying such EM signatures is also essential to associate future sources detected by LISA with their host galaxies \citep{AmaroSeoane:2023,Colpi:2024}.

GW recoils resulting from the merger of binary black holes can significantly influence the dynamical evolution of galaxy cores. These recoil velocities, which may \mbox{exceed} the escape speed of their host galaxies, can lead to the ejection of the remnant supermassive black hole (SMBH). Such events have profound implications for the cosmic SMBH population, the rate of subsequent mergers, and the coevolution of black holes with their host galaxies \citep{Volonteri:2007et,Sijacki:2010tk, Dunn:2020fgj}. This phenomenon is driven by the anisotropic emission of GWs \citep{peres1962classical,bekenstein1973gravitational}, and accurately predicting the recoil's magnitude requires full numerical relativity simulations due to the highly non-linear dynamics of the merger \citep{Redmount:1989, Campanelli:2004zw}. Early simulations revealed that, for certain binary configurations, these ``kicks'' could reach thousands of km\,s$^{-1}$ \citep{Campanelli:2007ew,Campanelli:2007cg, Gonzalez:2007hi,Lousto:2011kp}, \mbox{leading} to extensive observational searches for evidence of recoiling SMBHs.

These searches have targeted dynamical signatures, such as displaced galactic cores \citep{Volonteri:2007et,Loeb:2007wz,HolleyBockelmann:2007eh,sesana:2007zk,Blecha:2008mg,Shields:2008va}, as well as spectroscopic signatures, including large velocity offsets between broad and narrow emission lines in objects such as CID-42 \citep{Civano:2010es,Blecha:2012kx,Lanzuisi:2013fza} and 3C186 \citep{Chiaberge:2016eqf,Lousto:2017uav,Chiaberge:2018lkg}. For early reviews, see \citet{Komossa:2012cy}, \citet{Blecha:2015baa} and \citet{Sperhake:2014wpa}. 
Mergers of spinning equal-mass black holes produce the largest recoils. Specifically, the \textit{hang-up kick} configuration \citep{Lousto:2011kp}, where partially aligned spins delay the merger \citep{Campanelli:2006uy,Healy:2018swt}, due to strong spin-orbit coupling, and the asymmetric radiation is maximized, can produce recoils up to $5000$ km\,s$^{-1}$. 
The theoretical likelihood of this scenario involving different accretion models has been evaluated in \citet{Lousto:2012su}.
The observability of GWs from these extreme recoils has also been explored \citep{Gerosa:2016vip,CalderonBustillo:2018zuq,Lousto:2019lyf,Varma:2022pld,Wu:2025cun,Ranjan:2024wui}. While such vacuum predictions are well-established, the critical interplay between a \mbox{recoiling} black hole and its surrounding gaseous environment, responsible for \mbox{producing} the observable EM counterparts, has thus far been modeled with the recoil imposed rather than derived, on a fixed or Newtonian background \citep{Sijacki:2010tk, Ponce:2012, Meliani:2016rwn,Lippai:2008fx, franchini2024emission}. 
Relativistic magnetohydrodynamic treatments of the recoiling remnant have appeared very recently \citep{kim2026recoil}, but these considered a black hole boosted into a magnetically arrested circumbinary disk on a fixed background, neglecting the merger itself.

In this Letter, we present the first general \mbox{relativistic} magnetohydrodynamic (GRMHD) simulation of the gas dynamics around a rapidly recoiling SMBH, \mbox{following} a SMBBH in a \textit{hang-up kick} configuration, embedded within a relaxed circumbinary disk (CBD), through the entire dynamical evolution from late inspiral to merger and postmerger recoil.
During the inspiral phase, magnetically dominated relativistic jets are launched from the minidisks \mbox{surrounding} each spinning black hole through the Blandford--Znajek (BZ) mechanism \citep{BlandfordZnajek1977}. 
We observe that while the minidisks are tilted near the black holes according to their individual spin orientations, the jets rapidly realign with the binary's orbital angular momentum at larger distances.

The merger yields a spinning SMBH remnant that recoils at $\simeq 1032$ km\,s$^{-1}$ along the negative $z$-axis. The remnant spin is aligned with the orbital angular momentum of the binary, and therefore with that of the CBD, and the entire simulated disk remains gravitationally bound, comoving with the recoiling black hole.
Following coalescence, a relativistic jet is re-launched from the recoiling, newly spinning remnant, \mbox{maintaining} alignment with its final spin axis.

A central result is that coalescence brightens the system rather than dimming
it, in contrast to earlier simulations \citep[e.g.,][]{Krauth2023}: the \mbox{bolometric} luminosity rises by a factor $\sim\!2.2$ at nearly constant accretion rate, so that the emission is decoupled from the rest-mass supply through the horizon. The merger-driven spiral shocks responsible, identified for aligned spins by \citet{ennoggi2026merger}, persist throughout the recoil, and 
magnetized structures imprint a shared quasi-periodicity on the horizon flux, the Poynting flux, and the radiated output. The resulting ultraviolet \mbox{transient} is prompt on the timescale of the GW event, establishing recoiling remnants as postmerger counterparts for LISA.


\section{Numerical Techniques}

We perform a full GRMHD simulation of an equal-mass SMBBH in a {\it hang-up kick} configuration \citep{Lousto:2011kp}, with high, misaligned spins embedded in a magnetized thin CBD ($H/R \sim 0.1$) \citep{Noble:2012xz, LopezArmengol2021}. The \mbox{individual} spin vectors, each with magnitude \mbox{$\chi_i = s_i/m_i^2 = 0.8$}, are tilted by $45^\circ$ with respect to the binary orbital angular momentum. This setup yields in-plane spin components of equal magnitude but opposite sign, while the out-of-plane components share the same magnitude and sign, which, combining the hang-up \citep{Campanelli:2006uy} and superkick \citep{Campanelli:2007cg} effects, maximizes the recoil.

The initial conditions, including the relaxed, quasi-steady CBD state and the ``handoff'' procedure, are adopted from our previous work \citep{Ennoggi2025, ennoggi2026merger}. In that study, the CBD was initially threaded by a purely poloidal magnetic field that was larger (by a factor $ \sim 3.5$) than in the current follow-up simulation to ensure accurate resolution of the magnetorotational instability (MRI). While the CBD was evolved for $\sim\!165$ orbits to reach equilibrium, the present simulation \mbox{follows} the final $\sim\!40$ orbits of the late inspiral, merger, and the subsequent postmerger evolution.

Throughout the simulation, we use geometric units ($G=c=1$), so both time and distance are expressed in units of the binary total mass $M$. The gravitational radius is therefore $r_g \equiv GM/c^2 = M$. 

The simulation is performed within the \textsc{EinsteinToolkit} framework and carried out in two distinct stages. In the first stage, the initial CBD is evolved \mbox{using} the \texttt{SphericalNR} code~\citep{Mewes:2018szi, Mewes:2020vic, ji2023ameliorating} on a uniform, spherical-like mesh suited to the symmetries of the system. Once this initial \mbox{relaxation} phase is complete, the data are ``handed-off''~\citep{Ennoggi2025} and mapped onto a Cartesian grid, which serves as the starting point for the numerical binary evolution. In the second stage, spacetime initial data are constructed using a modified version of the \texttt{TwoPunctures} thorn~\citep{Ansorg:2004ds}, and the evolution is carried out with the Kranc-based \texttt{McLachlan} code via the implementation of the `moving puncture' approach~\citep{Campanelli:2005dd, Baker:2005vv}. The GRMHD equations are solved with the \texttt{IllinoisGRMHD} code~\citep{Etienne2015, Werneck2023}, with a simple ``target-entropy'' prescription to account for gas cooling via photon emission \citep{Ennoggi2025,Noble:2012xz}.

We employ an adaptive mesh refinement (AMR) grid structure provided by the \texttt{Carpet} driver~\citep{Schnetter-etal-03b}. Our Cartesian grid setup closely follows that of \citet{ennoggi2026merger}, featuring 14 \mbox{levels} of refinement and reaching a finest resolution of $\Delta x_f = M/128$. The finest level is located at radius $0.5 \ M$, ensuring that each apparent horizon is resolved by approximately 60--80 grid points across its diameter. Departing from \citet{ennoggi2026merger}, we move the second finest level from $1.5 \ M $ to $1\ M$.
Consistent with the predictions of \citet{Lousto:2012su} and \citet{Lousto:2019lyf}, the merger produces a remnant black hole with mass $m_f=0.93$ and spin magnitude $\chi_f=0.84$, aligned with the orbital angular momentum of the binary. 
Given the larger remnant radius, in the postmerger evolution, the hierarchy is reduced to 13 refinement levels, coarsening the finest resolution to $\Delta x_f = M/64$, while preserving the numerical accuracy.
A key result is the postmerger recoil velocity of the remnant black hole
$v_{\text{kick}} \simeq 1032$ km\,s$^{-1}$. The latter is measured by integrating the GW momentum fluxes, computed using the Weyl scalar $\psi_4$~\citep{Campanelli:1998jv, Lousto:2007mh} extrapolated to future null infinity using the techniques developed in~\citet{Nakano:2015pta}. 

\begin{figure*}[t]
  \centering
  \includegraphics[width=0.7\textwidth]
  {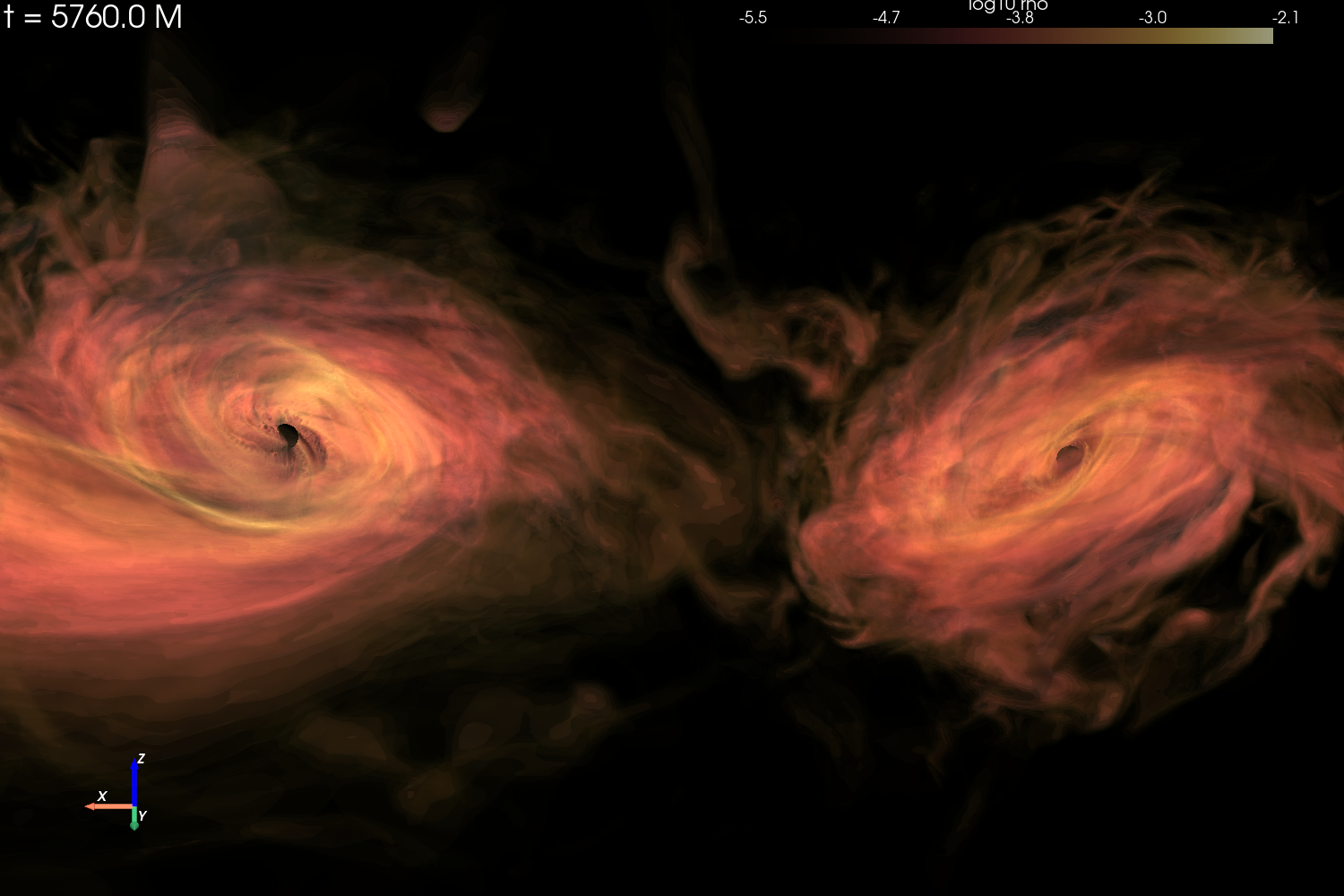}
  \caption{\justifying Volume rendering of the rest-mass density $\log_{10}\rho_b$ during the mid inspiral phase ($t=5760\,M$), extending $15\,M$ from the binary center of mass in each direction. The view is nearly edge-on, elevated  $25^\circ$ above the orbital ($x$--$y$) plane. Each black hole is surrounded by a strongly warped minidisk, tilted so as to follow the orientation of its individual spin, which is inclined by $45^\circ$ with respect to the orbital angular momentum. Due to the in-plane spin components being equal and opposite, the two minidisks are warped in opposite directions. The minidisks are continuously fed by accretion streams stripped from the inner edge of the circumbinary disk. 
  }
  \label{fig:volrender}
 
\end{figure*}

\begin{figure*}[t]
  \centering
  \includegraphics[width=\textwidth]{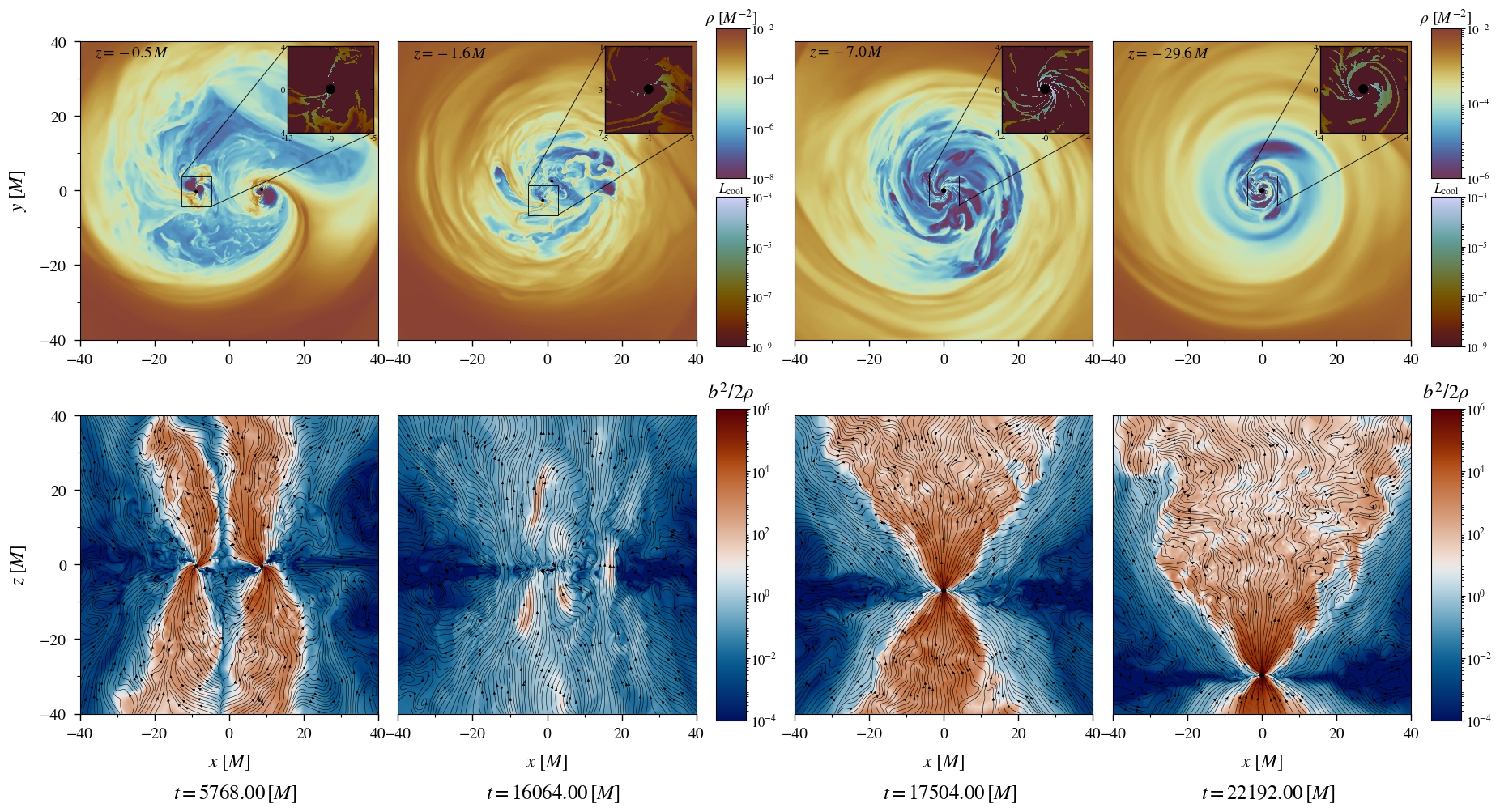}
  \caption{\justifying The top row shows the rest-mass density on ($x$--$y$) slices taken at the instantaneous height of the black holes; the progressive offset from $z=0$ in the postmerger columns is the displacement of the system along the recoil direction. The bottom row displays the magnetization parameter $b^2/2\rho$, on the corresponding meridional ($x$--$z$) slices, with magnetic field lines overplotted. The four snapshots correspond to successive stages of the evolution: mid inspiral ($t=5768\,M$), before merger ($t=16064\,M$), and two postmerger epochs during the recoil ($t=17504\,M$ and $t=22192\,M$). Zoomed insets in the top row show the cooling function $L_{\mathrm{cool}}$, with enhanced values tracing shock heating close to the individual black holes. Note that the two leftmost and two rightmost columns use different limits for $\rho$, as indicated by the two separate colorbars; the $L_{\mathrm{cool}}$ and $b^2/2\rho$ ranges are the same throughout.}
  \label{fig:2}
\end{figure*}

\begin{figure*}[t]
  \centering
  \includegraphics[width=\textwidth]{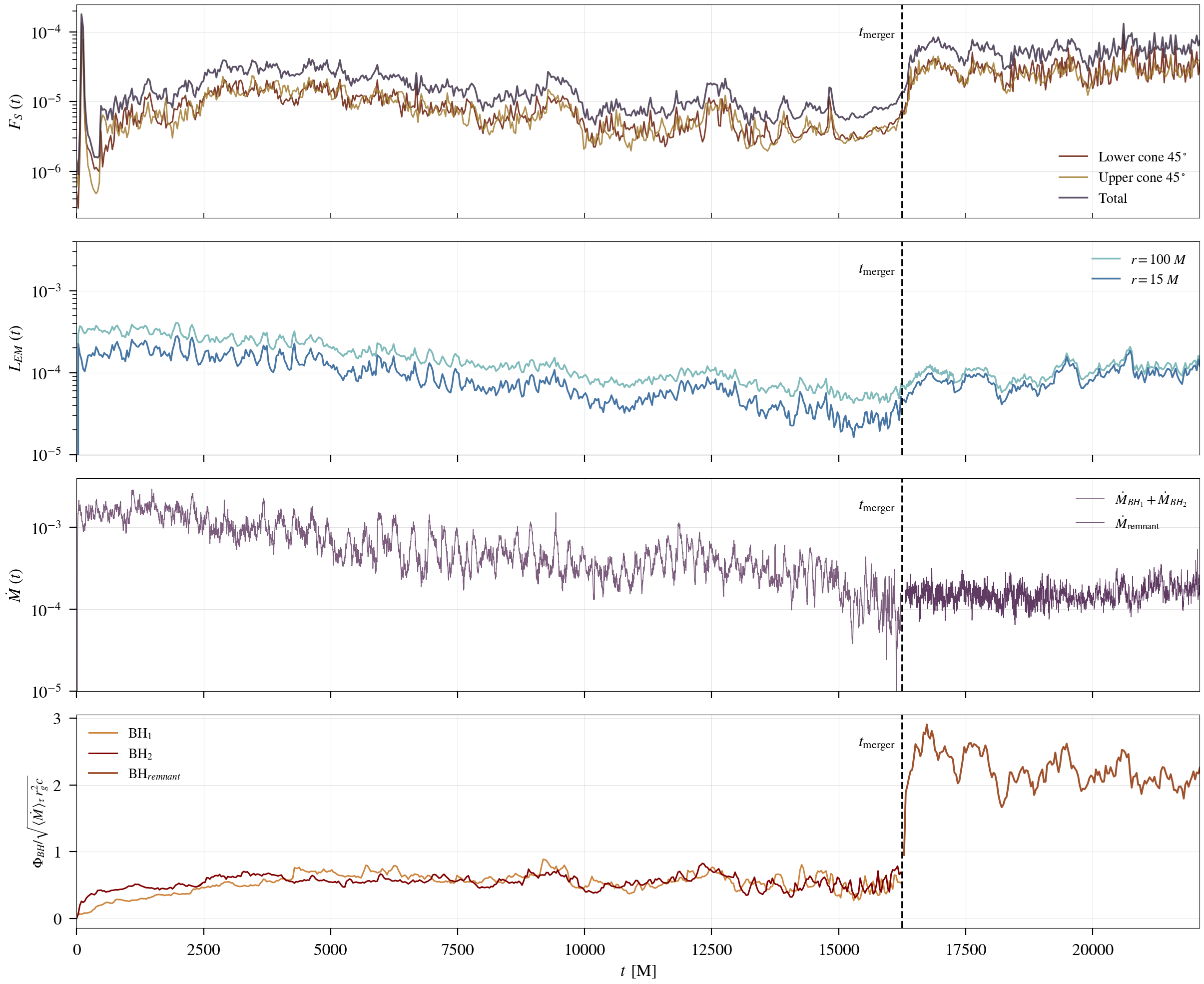}
  \caption{\justifying (Top) Poynting flux $F_S$, computed on $45^\circ$ aperture cones above and below the orbital plane at radius $r = 100\,M$ and shown for each cone separately and summed; the total is used throughout the text. (Second) Bolometric EM luminosity $L_{\rm EM}$, obtained by integrating the cooling rate $L_{\mathrm{cool}}$ over spheres of radii $r = 15\,M$ and $r = 100\,M$. (Third) Mass accretion rate onto the black hole horizons: the sum $\dot{M}_{\mathrm{BH}_1}+\dot{M}_{\mathrm{BH}_2}$ before merger, and $\dot{M}_{\rm remnant}$ after. (Bottom) Dimensionless horizon magnetic flux $\varphi_{\rm BH} \equiv \Phi_{\rm BH}/\sqrt{\langle\dot{M}\rangle_\tau r_g^2 c}$, with $\langle\dot{M}\rangle_\tau$ the accretion rate Gaussian-smoothed on $\tau = 1500\,M$, shown separately for $\mathrm{BH}_1$, $\mathrm{BH}_2$, and the postmerger remnant. The vertical dashed line marks the merger time.}
  \label{fig:3}
\end{figure*}

\begin{figure*}[t]
  \centering
  \begin{subfigure}{0.46\textwidth}
    \centering
    \includegraphics[width=\textwidth]{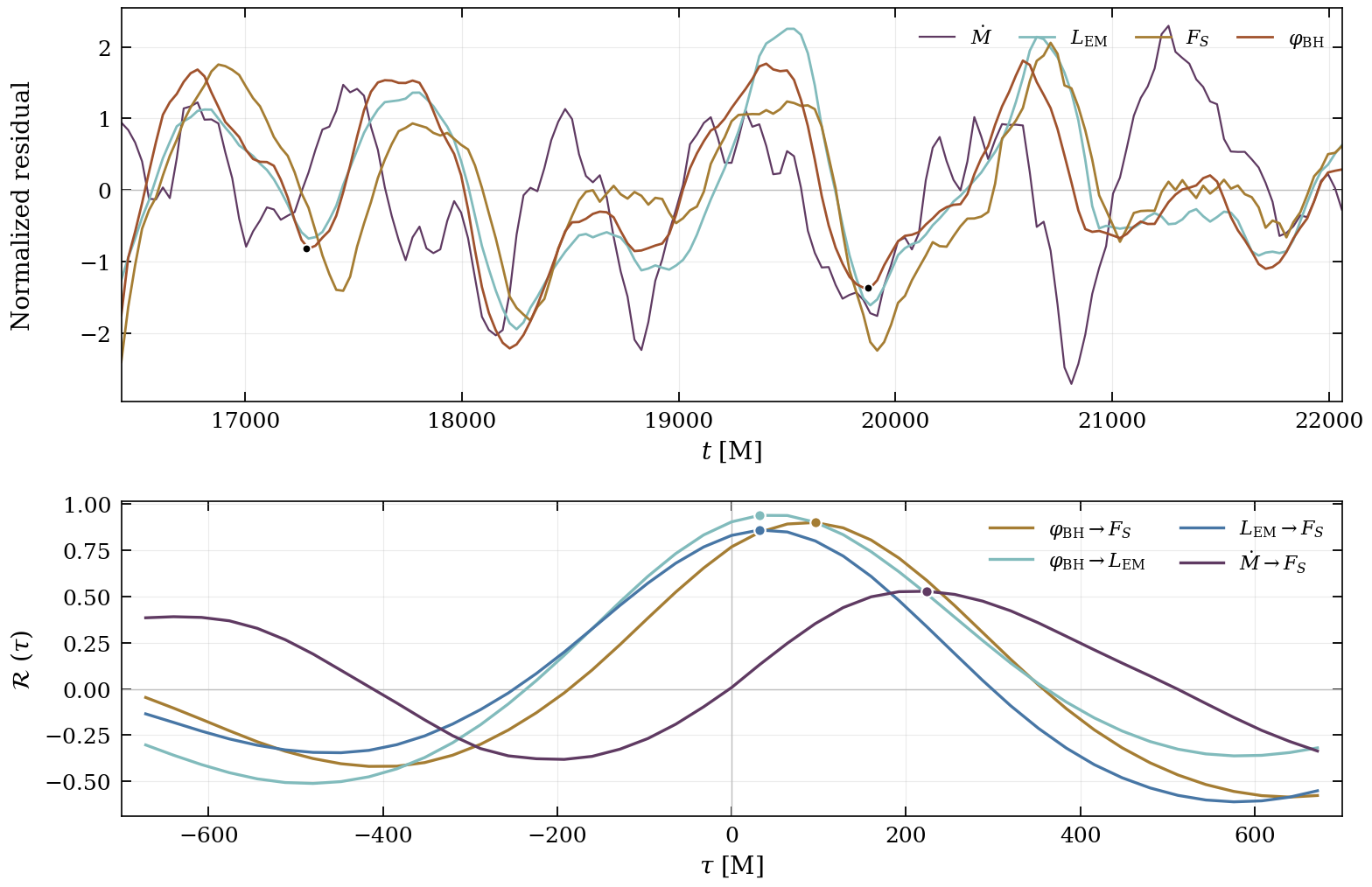}
    \caption{}
    \label{fig:corr}
  \end{subfigure}\hfill
  \begin{subfigure}{0.52\textwidth}
    \centering
    \includegraphics[width=\textwidth]{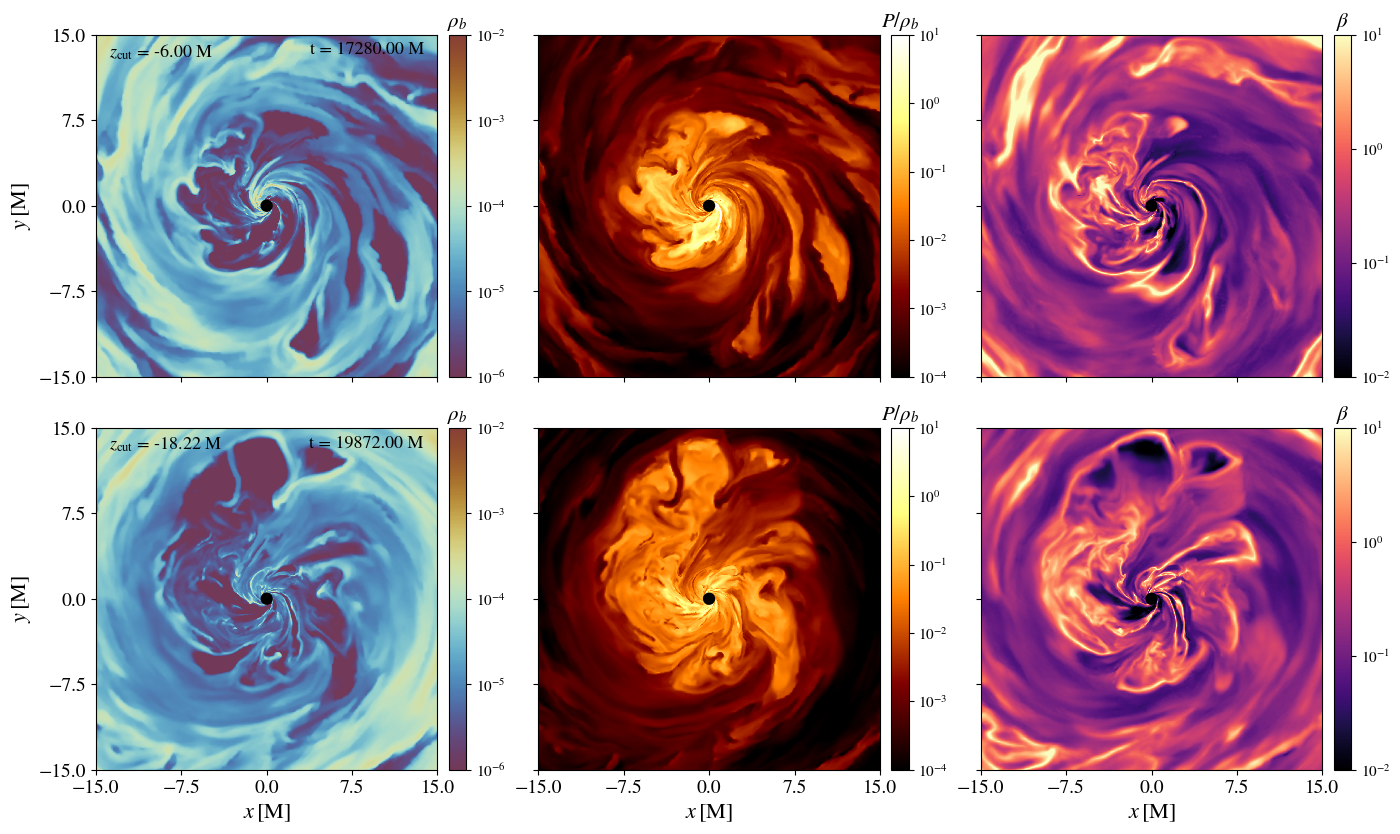}
    \caption{}
    \label{fig:slices}
  \end{subfigure}
  \caption{\justifying \textbf{a)} Postmerger residuals of the dimensionless horizon
  magnetic flux $\varphi_{\rm BH}$, the Poynting flux $F_S$, the bolometric luminosity
  $L_{\rm EM}$ at $r = 100\,M$, and the accretion rate $\dot{M}$, smoothed on
  $300\,M$ and with a $2000\,M$ secular trend removed (top), and the
  corresponding cross-correlation functions $\mathcal{R}(\tau)$,  where $X \rightarrow Y$ denotes $X$ leading $Y$ for $\tau > 0$ (bottom). \textbf{b)} Equatorial slices of the rest-mass density
  $\rho_b$, temperature $T \equiv P/\rho_b$, and plasma beta $\beta = 2P/b^2$ at
  the two minima of $\varphi_{\rm BH}$ marked with black circles in \textbf{a)}, at $t = 17280\,M$
  (top) and $t = 19872\,M$ (bottom).}
\end{figure*}

\section{Principal Results} 
\subsection{Inspiral and Merger}

We perform a fully relativistic simulation of the gas dynamics surrounding a recoiling SMBH formed from the merger of a SMBBH with misaligned spins. \mbox{Starting} from a binary separation of $a=20 \ M $, the simulation tracks the system through the final $\sim 40$ orbits of late inspiral, merger, and postmerger evolution. The initial gas distribution consists of a circumbinary disk \mbox{previously} evolved until equilibrium \citep{Ennoggi2025,ennoggi2026merger}. 
During the inspiral, each black hole retains its own minidisk, tilted by
$\simeq\!45^\circ$ so as to follow its individual spin axis and strongly warped (Fig.~\ref{fig:volrender}). Tidally stripped material is exchanged between the two minidisks across the L1 region, producing a quasi-periodic \textit{sloshing} \citep{Bowen:2016mci}.

Jets launched from the individual minidisks are tilted toward the spin axes of their individual black holes, with this alignment evident only near the jet base ($z \lesssim 5 \ M $), as shown in the first column of Fig.~\ref{fig:2}. At larger \mbox{distances}, however, the jets rapidly reorient, aligning with the orbital angular momentum of the binary.

As the binary approaches merger, the individual minidisks disappear and the
dual-jet structure is temporarily quenched, leaving a compact magnetized,
low-density structure that extends in the meridional plane at $r \simeq 20\,M$ (Fig.~\ref{fig:2}, second column). Following coalescence, the remnant SMBH receives a \textit{kick} of \mbox{$v_{\text{kick}} \simeq 1032$ km\,s$^{-1}$} directed along the negative $z$-axis. This phase is marked by an abrupt reduction in the binary's rest mass, with $ \sim 7\%$ of the total mass radiated in GWs, consistent with the empirical relation of \citet{TichyMarronetti2008}. Shortly after merger, a single, highly magnetized, relativistic jet is rapidly relaunched, preserving the large-scale alignment established during the pre-merger phase (Fig.~\ref{fig:2}, third column).

\subsection{Postmerger}
\label{sec:postmerger}

Theoretical estimates predict an outer gravitationally bound radius of $R_{\text{out}} \approx M/v_{\text{kick}}^2 \approx$ $8.45\times10^{4}\ M$ \citep{Loeb:2007wz}, within which the orbital speed exceeds the kick. Consequently, our simulation domain, with a half-edge length of $8192\ M$, focuses on the dynamically active core rather than the full theoretical extent of the disk. Nevertheless, this region contains the bulk of the CBD mass and captures the dominant physical processes of interest. The evolution of the loosely bound outer disk, and its delayed response to the recoil, is left to future large-scale simulations.

After merger, the inner boundary of the disk shrinks, the central cavity begins to fill and merger-driven spiral shocks persist throughout the disk during the postmerger phase and do not disrupt the coherent disk structure (see Fig.~\ref{fig:2}, top row, insets). 
Although we only evolved the postmerger for 
$\sim\!\dtpost\,M$, insufficient for the CBD to reach a new equilibrium, our results indicate that the disk is steadily progressing toward a relaxed state (Fig.~\ref{fig:2}, top row, fourth column). The $m=1$ orbital \textit{lump} \citep{Shi11,Noble:2012xz, Noble:2021vfg, LopezArmengol2021}, located at $2a \lesssim r \lesssim 3a$, remains visible \mbox{throughout} the merger and postmerger phase, as shown in both the density slices (Fig.~\ref{fig:2}, top row, third and fourth
columns) and the \mbox{azimuthal} mode amplitudes (see  Appendix~\ref{app:lump}). While the mode amplitude weakens, as expected, once the binary tidal torques cease to drive the gas, its \mbox{continued} survival offers a robust observational tracer of the \mbox{system's} dynamical history. 


To quantitatively characterize the evolving accretion dynamics, we track the mass accretion rate onto the black hole horizons alongside the EM luminosity, the Poynting flux, and the magnetic flux threading the horizons. These diagnostics are computed in the center of mass frame, following the techniques detailed in \citet{Ennoggi2025}. We adopt Heaviside--Lorentz units for the magnetic field throughout (a factor $4 \pi$ is absorbed in the definition of the magnetic energy density), so that the horizon magnetic flux is defined as 

\begin{equation}
  \Phi_{\rm BH} = \frac{1}{2}
  \int_{0}^{2\pi}\!\!\int_{0}^{\pi} |B^r|\, \sqrt{-g}\;
  \mathrm{d}\theta\, \mathrm{d}\phi ,
  \label{eq:phibh}
\end{equation}

and its dimensionless counterpart as $\varphi_{\rm BH} \equiv
\Phi_{\rm BH}/\sqrt{\langle\dot{M}\rangle_\tau r_g^2 c}$. In this convention, $\varphi_{\rm BH}$ saturates at $\simeq\!15$ for magnetically arrested flows
\citep{tchekhovskoy2011efficient}.
As shown in Fig.~\ref{fig:3}, through the late inspiral, the Poynting flux declines steadily, reaching a minimum of $\simeq\!1\times10^{-5}$ shortly before coalescence as the minidisks are depleted and the dual-jet structure is quenched, while at merger it rises sharply, settling at $\simeq\!4\times10^{-5}$. This strengthening tracks the magnetic flux threading the horizon (Fig.~\ref{fig:3}, bottom panel), as expected for BZ-powered emission.  During the inspiral, each black hole reaches $\varphi_{\rm BH} \simeq 0.5$--$0.8$; at coalescence, the remnant
peaks at $\varphi_{\rm BH} \simeq 2.9$ and thereafter \mbox{fluctuates} between
$\simeq\!1.9$ and $\simeq\!2.5$. Nevertheless, the flow remains well below the magnetically arrested disk (MAD) threshold throughout the simulation.

As a result of the merger-driven reconfiguration of the gas flow and the associated shock dissipation, 
the \mbox{bolometric} EM luminosity increases by a factor of $\sim 2.2$, from $4.53 \times 10^{-5}$ to $1.03 \times 10^{-4}$ ($r \leq 100 \ M$), after merger while the mass accretion rate declines by only $\simeq\!15\%$ (Fig.~\ref{fig:3}). 
The enhanced emission is concentrated within $r \le 15\,M$, as evidenced by the near coincidence of the $r = 15\,M$ and $r = 100\,M$ postmerger luminosity curves. Consequently, the aftermath of the merger phase emerges as the most luminous component of the \mbox{system}, providing a robust and observable EM \mbox{counterpart} for LISA-band gravitational wave events \citep{AmaroSeoane:2023}.


We find that the tight coupling between the photon and the Poynting emission, identified during the inspiral phase of the aligned-spin binary configuration \citep{ennoggi2026merger}, persists throughout the postmerger and extends to the horizon magnetic flux, resulting in a shared low-frequency, quasi-periodic modulation ($P \simeq 10^{3}\,M$ over the few cycles we resolve). In order to quantify this coupling, we compute the Pearson cross-correlation $\mathcal{R}(\tau)$ as a function of the lag, since the four diagnostics are extracted on different surfaces. After subtracting a $2000\,M$ secular trend and smoothing over $300\,M$,  we show that the residual modulations of $\varphi_{\rm BH}$, $F_S$, and $L_{\rm EM}$ follow one another closely, $\mathcal{R} \simeq 0.9$, while $\dot{M}$ follows only weakly and much later ($\mathcal{R} \sim 0.55$, at $\tau \simeq 2\times10^{2}\,M$). This implies that the jet and the radiative output are modulated coherently, and they are not in step with the instantaneous rest-mass supply (Fig.~\ref{fig:corr}). 

The minima of $\varphi_{\rm BH}$ are associated with the ejection of magnetic flux bundles close to the horizon, as seen in magnetically arrested flows around single black holes \citep{tchekhovskoy2011efficient,mckinney2012general,ripperda2022black} and in magnetically arrested minidisks in CBD accretion \citep{manikantan2025magnetically}. Two such minima, separated by two to three modulation cycles, are marked in Fig.~\ref{fig:corr} and shown in Fig.~\ref{fig:slices}. At the earlier minimum the magnetized, relativistically hot material remains confined to a few narrow lanes within $r \lesssim 7\,M$, resembling the ``miniflares'' of \citet{ripperda2022black}, in which flux is expelled close to the horizon without evacuating the disk. At the later one, low-$\beta$, hot ($T \sim 0.1$--$1$) material extends beyond $r \simeq 10\,M$, suggesting a
large-scale expulsion.


Here, however, the flow is globally sub-MAD, so any such events must be driven by locally, rather than globally, saturated magnetic pressure generated by merger-driven shock compression. We also note that plasmoid-mediated reconnection would require a substantially higher resolution and is therefore not captured.

The abrupt mass loss, together with the recoil imparted to the remnant, is expected to leave the bound gas on eccentric orbits with $e \approx \Delta M/M_{\rm tot}$
\citep{ONeill2009,Lippai:2008fx,SchnittmanKrolik2008}. For our configuration this expectation is marginal: the criterion of \citet{ONeill2009}, by which mass-loss
driven shocks require $\Delta M/M_{\rm tot} > H/R$, gives
$\Delta M/M_{\rm tot} \simeq 0.07$ against $H/R \approx 0.1$. We therefore measure the disk eccentricity across multiple radii, and find $e \lesssim 0.05$ in the annulus at $r \simeq 40$--$60\,M$, falling below $0.01$ beyond $r \simeq 80\,M$, with no discontinuity across coalescence (see  Appendix~\ref{app:lump} for details). The perturbation therefore remains subsonic and is smoothed by pressure, in contrast with the stronger responses
predicted by ballistic and thin-disk models
\citep{Corrales:2010,Lippai:2008fx,Shields:2008va,SchnittmanKrolik2008,Rossi:2010}.

It is useful to convert these diagnostics into physical units. Because the gas is neither self-gravitating nor radiatively back-reacting, the density normalization is arbitrary.
We write $M_8 \equiv M/10^{8}\,M_\odot$ and $\dot{m}_{0.1} \equiv \dot{m}/0.1$, with $\dot{m} \equiv \dot{M}/\dot{M}_{\rm Edd}$ and $\dot{M}_{\rm Edd} \equiv L_{\rm Edd}/(0.1\,c^{2})$. 

The radiative efficiency measured from Fig.~\ref{fig:3},
$\varepsilon \equiv L_{\rm EM}(r \le 100\,M)/\dot{M}c^{2}$, rises from
$\varepsilon_{\rm insp} \simeq \radeffins$ during the late inspiral to
$\varepsilon_{\rm post} \simeq \radeffpost$ after merger. Averaged over matched $2000\,M$ windows either side of coalescence, $L_{\rm EM}$ increases by a factor $2.2$ while $\dot{M}$ declines by only $\simeq\!15\%$, so that $\varepsilon$ rises by
a factor $\simeq\!2.6$. It is this rise at nearly constant accretion
rate, rather than the magnitude of $\varepsilon$ itself, that identifies the merger as the energy source: the luminosity is decoupled from the rest-mass supply through the horizon. 
Previous aligned-spin \mbox{simulations} \citep{ennoggi2026merger} reach $\varepsilon \sim 1$--$2$ in the same phase.

In physical units this gives
\begin{equation}
  L_{\rm post} \simeq \frac{\varepsilon_{\rm post}}{0.1}\,\dot{m}\,L_{\rm Edd}
  \simeq 8\times10^{45}\,\dot{m}_{0.1}\, M_8 \;\; {\rm erg\,s^{-1}},
  \label{eq:Lpost}
\end{equation}

a factor $\sim\!2.2$ above the late-inspiral level. Since the emission originates within $r \lesssim 15\,M$, the corresponding effective temperature is $T_{\rm eff} \simeq 1.2\times10^{5}\,\dot{m}_{0.1}^{1/4}M_8^{-1/4}$ K, i.e.\ $kT_{\rm eff} \simeq 10\,\dot{m}_{0.1}^{1/4}M_8^{-1/4}$~eV: an
ultraviolet transient that hardens toward lower masses, from
$kT_{\rm eff} \simeq 6$~eV at $10^{9}\,M_\odot$ to $\simeq\!30$~eV for a
$10^{6}\,M_\odot$ remnant. With $t_g \equiv GM/c^{3} \simeq 4.9\times10^{2}\,M_8$ s, the luminosity jump is unresolved on scales of $\sim\!10^{2}\,M \simeq 0.6\,M_8$ days and the shocked, luminous state persists for at least the $\dtpost\,M \simeq 33\,M_8$ days we evolve. For a LISA-band remnant of $10^{6}\,M_\odot$ the rise is $\lesssim\!10$ minutes and the enhanced state lasts $\gtrsim\!8$ hours; at $10^{9}\,M_\odot$ these become $\lesssim\!6$ days and $\gtrsim\!11$ months. The counterpart is thus prompt on the timescale of a GW event.



\section{Discussion}


The prompt luminosity brightening we find differs qualitatively from the Newtonian 2D
hydrodynamic SMBBH simulations of \citet{Krauth2023}, which report dimming
before and after merger. With non-linear hydrodynamics and magnetic effects
included, shock heating replaces minidisk-driven dissipation as the dominant
emission source once the minidisks are destroyed during the late inspiral, and
the shocks persist throughout the recoil. Despite the extreme postmerger disturbance, the disk preserves its structural coherence. This turbulent, shock-dominated response is captured only by fully relativistic simulations.



The correlated modulation of the horizon magnetic flux, the Poynting flux and the
thermal output is tied to the merger rather than to the accretion state: the flow is globally sub-MAD, so the magnetic pressure driving it must
saturate locally rather than globally, plausibly through shock compression at
coalescence. The coupling persists for the full postmerger evolution we follow, with the jet and the radiated output tracking one another but not the instantaneous accretion rate, behavior that ordinary accretion variability does not produce.

The relativistic jet launched from the recoiling SMBH remains aligned with the binary's orbital angular momentum, whose direction is preserved through coalescence and the subsequent recoil. 
This is consistent with  simulations of spinning binaries in magnetized gas clouds \citep{cattorini2022misaligned} and thicker disks \citep{ruiz2023general}, as well as with studies of single black holes in comparable environments \citep{ressler2021magnetically,kelly2021electromagnetic}. By contrast, more generic precessing binaries are expected to exhibit more complex dynamics, potentially giving rise to weaker, more intermittent, and less well-structured jets near merger \citep{Ressler2024}.

Our results confirm the qualitative theoretical \mbox{prediction} that a recoiling remnant retains its accretion disk \citep{Loeb:2007wz}, thereby preserving the conditions necessary for continued EM emission. 
Because this EM signature is fundamentally shaped by the violent gas dynamics at coalescence, simplified `boosted disk' models initialized in equilibrium artificially suppress the powerful spiral shocks inherited from the merger. Capturing these features is essential for predicting the prompt luminosity enhancement and for distinguishing these signatures from standard active galactic nuclei activity.

Interestingly, this scenario aligns closely with the \mbox{leading} recoiling SMBH candidate, 3C186 \citep{Chiaberge:2016eqf,Chiaberge:2018lkg,Morishita:2022aro,Castignani:2022ipv}, for which the inferred kick velocity is $\sim 1300$ km\,s$^{-1}$ \citep{Chiaberge:2016eqf, Lousto:2017uav, Boschini:2024tls}, comparable to the
$\simeq\!1032$ km\,s$^{-1}$ we obtain. 




Such sources are observed $\gtrsim 10^6$ yr postmerger, by which the accretion disk has fully equilibrated. Our simulation, instead, resolves the prompt, violent phase which lasts $\sim\!11$ months for a $10^{9}\,M_\odot$ remnant and $\gtrsim\!8$ hours for a $10^{6}\,M_\odot$ LISA-band remnant (Sec.~\ref{sec:postmerger}). It therefore constrains the engine that powered such sources rather than their present-day appearance, and predicts an EM counterpart for LISA mergers.



\section{Conclusion}

In this Letter, we presented the first self-consistent GRMHD simulation of a misaligned SMBBH  merger in a relaxed circumbinary disk, followed from late inspiral through coalescence to the recoil of the remnant. Contrary to the dimming expected from mass-loss arguments
and reported by earlier Newtonian models, we find that the merger brightens the
system by a factor $\sim\!2.2$ while the mass accretion rate declines by only $\simeq\!15\%$: the radiative efficiency rises from $\varepsilon\simeq 0.23$ to $\simeq\!0.6$ and, being the emission decoupled from the rest-mass supply through the horizon, it is
coalescence, not accretion, that powers it. The horizon magnetic flux and the Poynting flux both rise sharply at merger, after the dual-jet structure is quenched during the late inspiral, while the flow remains globally sub-MAD throughout.

We also find that the jet launched by the recoiling SMBH preserves the large-scale \mbox{alignment} established prior to merger with the orbital angular momentum of the binary. Despite the strong dynamical perturbations associated with coalescence and a recoil of $\simeq 1032$ km\,s$^{-1}$, the accretion disk within our computational domain remains gravitationally bound and preserves its structural coherence, and the $m=1$ lump survives as a tracer of the system's dynamical history. Shock-compressed magnetic structures in the inner flow intermittently drive correlated modulations of the horizon magnetic flux, the Poynting flux, and the thermal output ($\mathcal{R}\simeq0.9$), with the accretion rate responding only weakly and at a longer delay. In physical units, the postmerger state radiates at
$\sim\!8\times10^{45}\,\dot{m}_{0.1}M_8\,{\rm erg\,s^{-1}}$: an ultraviolet
transient that switches on at coalescence. For a $10^{6}\,M_\odot$ LISA-band
remnant it rises within $\lesssim\!10$ minutes and stays enhanced for
$\gtrsim\!8$ hours; at $10^{9}\,M_\odot$ these become $\lesssim\!6$ days and
$\gtrsim\!11$ months. The spectral peak is
largely inaccessible to extragalactic observation, but the Rayleigh--Jeans
tail extends into the optical, where a transient of this duration is within
reach of wide-field surveys and can be localized to its host galaxy.



\begin{acknowledgments}
This work was supported by a NASA Theory and Computational Astrophysics Network (TCAN) grant, 80NSSC24K0100. M.C., C.O.L., and Y.Z. acknowledge additional support from NSF awards AST-2009330, AST-1516150, AST-2319326, PHY-2110338, PHY-1707946, PHY-2207920, PHY-2513442, PHY-2409706, and OAC-2411068.

We thank Giuseppe Ficarra
for independently verifying our recoil kick velocity calculation. We are also grateful to Michail Chabanov, Luciano Combi, Jay V. Kalinani, Julian Krolik, Scott Noble, Frederic A. Rasio, Liwei Ji, Vassilios Mewes, and Jeremy Schnittman for valuable discussions and comments.

The authors acknowledge the Texas Advanced Computing Center (TACC) for providing computational resources on the Frontera supercomputer through allocations PHY-20010 and AST-20021. This research also used resources at the Rochester Institute of Technology (RIT), including the BlueSky,
Green Prairies, and Lagoon clusters, which were acquired with support from NSF grants PHY-2018420, PHY-0722703, PHY-1229173, and PHY-1726215.

Claude (Anthropic) was used to assist with the final manuscript revisions and consistency checking; all scientific content is the authors' own.
\end{acknowledgments}


\appendix

\section{Evolution of the lump and disk eccentricity}
\label{app:lump}

\begin{figure*}[h!]
  \centering
  \begin{subfigure}{0.35\textwidth}
    \centering
    \includegraphics[width=\textwidth]{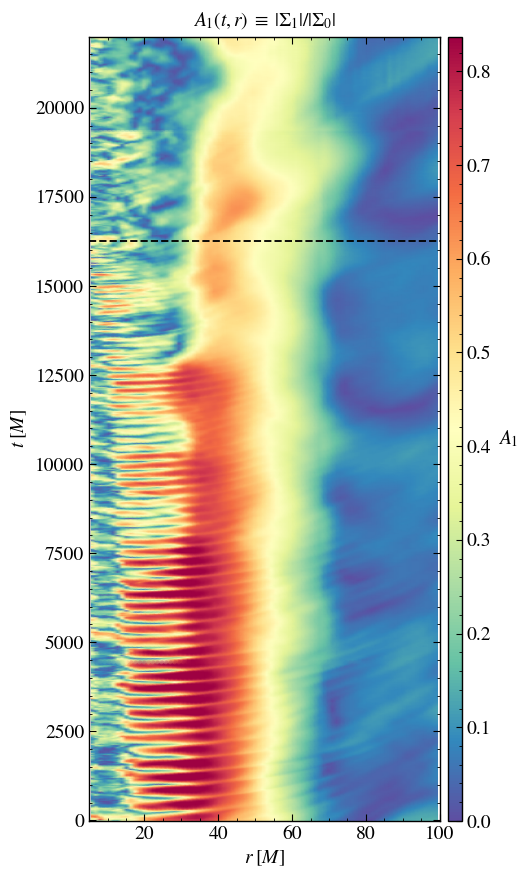}
  \end{subfigure}
  \hspace{0.01\textwidth}
  \begin{subfigure}{0.35\textwidth}
    \centering
    \includegraphics[width=\textwidth]{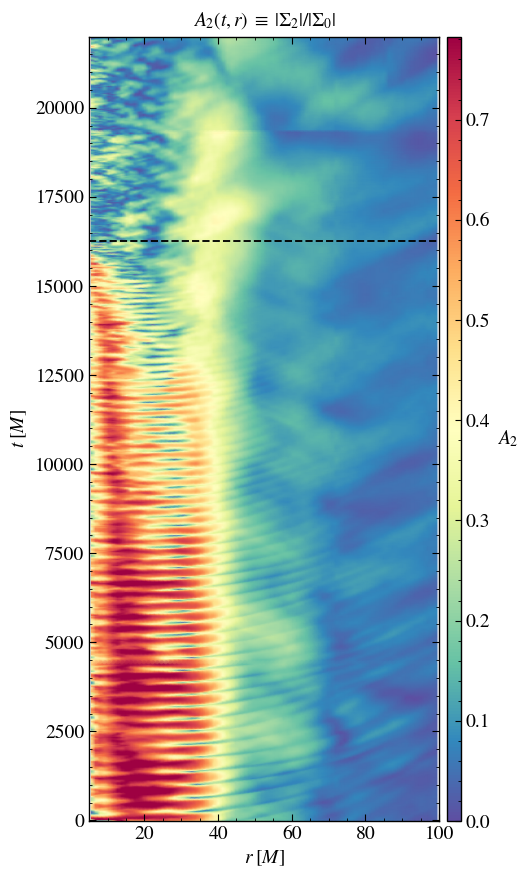}
  \end{subfigure}
  \caption{\justifying Normalized azimuthal Fourier amplitudes of the surface density, $A_1(t,r) \equiv |\Sigma_1|/|\Sigma_0|$ (left) and $A_2(t,r) \equiv |\Sigma_2|/|\Sigma_0|$ (right), as functions of radius and time. The black dashed line marks the merger time.}
  \label{fig:ab_spectrograms}
\end{figure*}

The $m=1$ overdensity, or \textit{lump}, forms through a feedback cycle at the inner edge of the circumbinary disk: gas peeled into streams by the binary is partly flung back outward, having gained angular momentum, and shocks against the cavity wall, seeding an $m=1$ asymmetry that in turn modulates the subsequent stream production \citep{Shi11, Noble:2012xz}. In order to track the evolution of its amplitude, we decompose the vertically integrated density $\Sigma(t,r,\phi)$ into azimuthal Fourier modes,

\begin{equation}
  \Sigma_m(t,r) = \frac{1}{2\pi}\int_0^{2\pi} \Sigma(t,r,\phi)\,e^{-im\phi}\,d\phi ,
\end{equation}

and analyze the normalized amplitudes $A_m(t,r) \equiv |\Sigma_m|/|\Sigma_0|$ for $m=1$ and $m=2$ (see Fig.~\ref{fig:ab_spectrograms}).

\begin{figure*}[h!]
  \centering
  \includegraphics[width=0.7\textwidth]{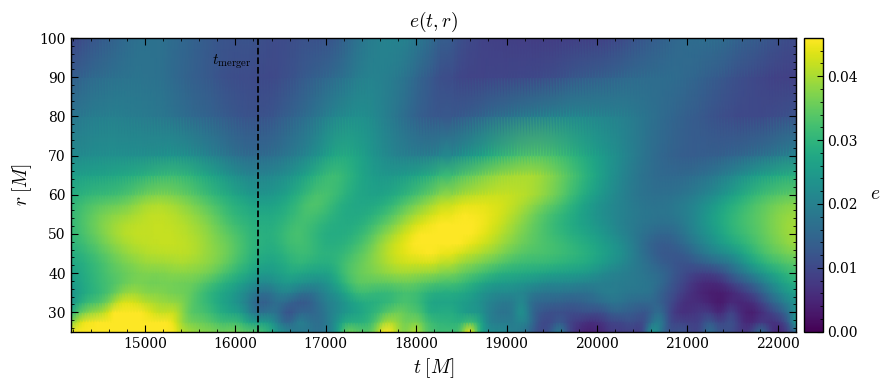}
  \caption{\justifying Disk eccentricity $e(t,r)$, computed on cylindrical shells over $25 \le r \le 100\,M$ and integrated to $|z| = 30\,M$, as a function of radius and time. A Gaussian smoothing of $2\,M$ in radius and $64\,M$ in time is applied. The black dashed line marks the merger.}
  \label{fig:eccentricity}
\end{figure*}

The resulting overdensity forms at $r \approx 2.4a \approx 48\,M$ for our initial separation $a = 20\,M$, and orbits close to the local Keplerian rate, $\Omega_{\rm lump} \approx 0.28\,\Omega_{\rm bin}$. Throughout the inspiral, both amplitudes are large, $A_1 \simeq 0.6$--$0.8$ and $A_2 \simeq 0.7$--$0.8$, though they peak at different radii: $A_2$ is strongest at $8\,M \lesssim r \lesssim 35\,M$, where the quadrupolar response to the binary potential dominates, while $A_1$ peaks slightly farther out, at $15\,M \lesssim r \lesssim 40\,M$. Both are modulated at the beat frequency between the binary and the lump ($f_{\rm beat} = f_{\rm orb}- f_{\rm lump}$). The modulation weakens well before coalescence, at $t \simeq 1.2$--$1.3\times10^{4}\,M$
for $m=1$, as the tidal truncation radius
$r_{\rm t} \propto a$ shrinks toward the ISCO and the minidisks are depleted, whereas the $m=2$ response, forced directly by the binary quadrupole, persists until merger. After coalescence, a coherent $m=1$ feature survives at $35\,M \lesssim r \lesssim 55\,M$
with $A_1 \simeq 0.6$--$0.7$ and decays slowly, while $A_2$ falls to $\simeq 0.3$--$0.4$ over the same annulus, since it is directly tied to the binary torque.

The abrupt loss of $\sim\!7\%$ of the system's mass-energy at coalescence, together with the recoil imparted to the remnant, is expected, on ballistic grounds, to leave the bound gas on eccentric orbits with $e \approx \Delta M / M_{\rm tot}$ \citep{ONeill2009,Lippai:2008fx,SchnittmanKrolik2008}. In order to test this, we evaluate the eccentricity of the surviving disk on cylindrical shells, following the equations in  \citet{Noble:2021vfg} and \citet{Shi11}. 

Figure~\ref{fig:eccentricity} shows that the eccentricity is everywhere small,
$e \lesssim 0.05$, concentrated in an annulus at $r \simeq 40$--$60\,M$ that
persists across the merger without discontinuity, and falls below $0.01$ beyond
$r \simeq 80\,M$. We further find the eccentricity phase to be coherent across
$35\,M \lesssim r \lesssim 100\,M$ and continuous across coalescence,
confirming that the residual eccentricity traces the surviving overdensity
rather than the mass loss.


\bibliographystyle{aasjournalv7}
\bibliography{references2,references,references3}  

\end{document}